\documentclass[prx,superscriptaddress,citeautoscript,twocolumn]{revtex4-2}

\usepackage{graphicx}
\usepackage{amsmath,amssymb}
\usepackage{color}
\usepackage{textcomp}
\usepackage{upgreek}
\usepackage{latexsym}
\usepackage{soul}
\usepackage{hyperref}
\hypersetup{
	colorlinks=true,
	linkcolor=red,
	filecolor=blue,
	urlcolor=blue,
	citecolor=blue,
}

\begin{document}

\title{Flexible terahertz metasurface absorbers empowered by bound states in the continuum}

%


\author{Guizhen Xu}
\affiliation{Department of Electrical and Electronic Engineering, State Key Laboratory of Optical Fiber and Cable Manufacture Technology, Southern University of Science and Technology, Shenzhen 518055, China}

\author{Zhanqiang Xue}
\affiliation{Department of Electrical and Electronic Engineering, State Key Laboratory of Optical Fiber and Cable Manufacture Technology, Southern University of Science and Technology, Shenzhen 518055, China}

\author{Junxing Fan}
\affiliation{Department of Electrical and Electronic Engineering, State Key Laboratory of Optical Fiber and Cable Manufacture Technology, Southern University of Science and Technology, Shenzhen 518055, China}

\author{Dan Lu}
\affiliation{Department of Electrical and Electronic Engineering, State Key Laboratory of Optical Fiber and Cable Manufacture Technology, Southern University of Science and Technology, Shenzhen 518055, China}

\author{Hongyang Xing}
\affiliation{Department of Electrical and Electronic Engineering, State Key Laboratory of Optical Fiber and Cable Manufacture Technology, Southern University of Science and Technology, Shenzhen 518055, China}

\author{Perry Ping Shum}
\affiliation{Department of Electrical and Electronic Engineering, State Key Laboratory of Optical Fiber and Cable Manufacture Technology, Southern University of Science and Technology, Shenzhen 518055, China}
\affiliation{Guangdong Key Laboratory of Integrated Optoelectronics Intellisense, Southern University of Science and Technology, Shenzhen 518055, China}

\author{Longqing Cong}
\email{\text{\textcolor{black}{Corresponding author:}} conglq@sustech.edu.cn}
\affiliation{Department of Electrical and Electronic Engineering, State Key Laboratory of Optical Fiber and Cable Manufacture Technology, Southern University of Science and Technology, Shenzhen 518055, China}
\affiliation{Guangdong Key Laboratory of Integrated Optoelectronics Intellisense, Southern University of Science and Technology, Shenzhen 518055, China}

\begin{abstract}
  \noindent
  {\bf Abstract:} Terahertz absorbers are crucial to the cutting-edge techniques in the next-generation wireless communications, imaging, sensing, and radar stealth, as they fundamentally determine the performance of detectors and cloaking capabilities. It has long been a pressing task to find absorbers with customizable performance that can adapt to various environments with low cost and great flexibility. Here, we demonstrate perfect absorption empowered by bound states in the continuum (BICs) allowing for the tailoring of absorption coefficient, bandwidth, and field of view. The one-port absorbers are interpreted using temporal coupled-mode theory highlighting the dominant role of BICs in the far-field radiation properties. Through a thorough investigation of BICs from the perspective of lattice symmetry, we unravel the radiation features of three BIC modes using both multipolar and topological analysis. The versatile radiation capabilities of BICs provide ample freedom to meet specific requirements of absorbers, including tunable bandwidth, stable performance in a large field of view, and multi-band absorption using a thin and flexible film without extreme geometric demands. Our findings offer a systematic approach to developing optoelectronic devices and demonstrate the significant potential of BICs for optical and photonic applications which will stimulate further studies on terahertz photonics and metasurfaces.
  \end{abstract}

\maketitle

\section{Introduction}

Terahertz radiation, characterized by its great penetration capability and non-ionizing property, is attracting increasing attention in an interdisciplinary field such as biology, security screening, medical imaging, and semiconductor industry\textsuperscript{\cite{1}}. Nowadays, a wide range of applications are becoming reality with terahertz waves including the next-generation wireless communications, radar, imaging, and sensing\textsuperscript{\cite{2,3,4}}. However, lack of high-performance terahertz absorbers has hampered the proliferation of these promising applications which are a key component to capture weak terahertz signals for detectors and reduce scattering cross section for security communications and cloaking of aircraft. In practice, an ideal terahertz absorber should possess a stable spectral response to incidence of various angles and a controllable bandwidth in addition to a large absorption coefficient as well as a thin and flexible film applicable for fluctuating surfaces. It has always been a pressing task to look for solutions of absorbers that could meet one or more of those requirements for practical usage. Recently, excellent absorption performance was reported with an ultrathin Ti$_3$C$_2$T$_x$ MXene film covering the whole terahertz range, but below a theoretically limited absorption coefficient of $50\%$\textsuperscript{\cite{5}}. Despite the significance of such MXene film, the intrinsic chemical instability deteriorates the absorption performance over time which is prone to oxidation. In addition, it is difficult to control the size and defects of MXene flakes that would significantly impact its practical performance. 

\begin{figure*}
	\centerline{\includegraphics[width=0.86\textwidth]{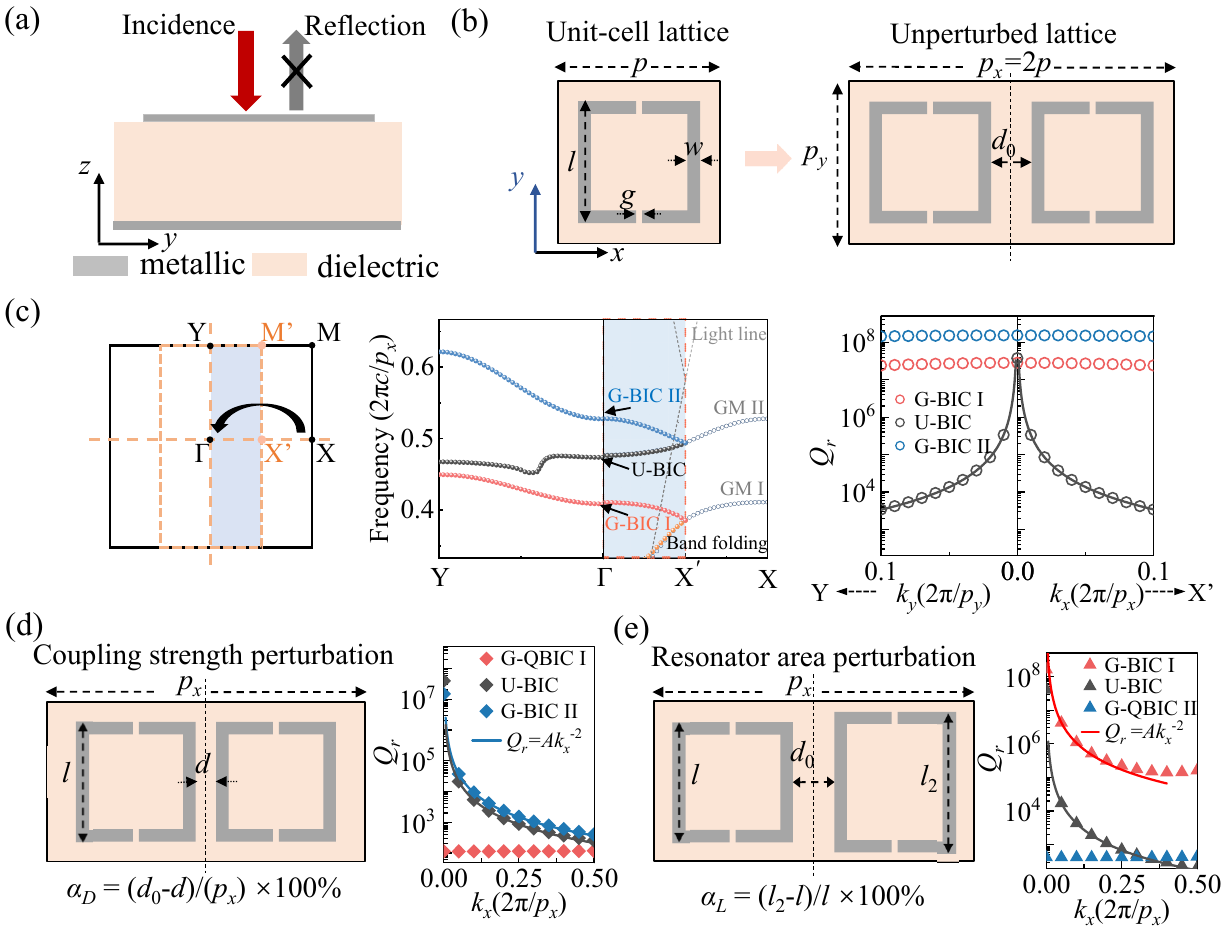}}
	\caption{{\bf Schematic diagram of MIM metasurfaces and band folding process.} (a) One-port system composed of a MIM cavity. (b) A unit cell in a square lattice and a supercell with two DSRRs in a rectangular lattice. (c)Illustration of Brillouin zone for rectangular lattice (dashed lines) and square lattice (solid lines, left), band structure diagram before and after band folding (middle), and the corresponding radiative quality factors $Q_r$ of U-BIC and G-BICs in the momentum space after band folding (right). The calculation was performed with parameters when $p$ = 100 $\mu$m, $l$ = 75 $\mu$m, $w$ = 8 $\mu$m, $g$ = 3 $\mu$m, and $d_0$ = 25 $\mu$m. (d) and (e) Schematic diagram of rectangular lattice by perturbing coupling strength ($d$ = 2 $\mu$m, $\alpha_D$ = 11.5\%) and resonator area ($l_2$ = 80 $\mu$m, $\alpha_L$ = 6.7\%) with their evolution of radiative $Q_r$ versus $k_x$ for G-BIC I (red), U-BIC (grey) and G-BIC II (blue). The solid lines are fitting curves with $Q_r \propto k_x^{-2}$. Perfect electric conductor (PEC) and lossfree substrate were used in simulations to calculate the eigenvalues.}
	\label{Fig1}
\end{figure*}

Metasurfaces offer a promising platform to probe the underlying mechanism of accessing an ideal absorber by accurately tailoring the geometries, integrating to various types of materials, and exploring interesting physics\textsuperscript{\cite{6,7,8,9}}. With great design flexibility, metasurfaces have realized terahertz absorption with unity coefficient, thin thickness, and flexible features\textsuperscript{\cite{10,11}} that are chemically stable in harsh environment and would be particularly useful in the applications of radar stealth and signal shielding in terahertz regime\textsuperscript{\cite{12,13}}. Typical configuration of a metamaterial absorber consists of three layers: a structured metasurface, a dielectric layer and a metallic ground (Fig. \ref{Fig1}(a))\textsuperscript{\cite{14}}. Based on this type of configuration, absorption with a high coefficient covering a wide incident angle and/or a broad frequency band was reported\textsuperscript{\cite{15,16,17,18}}. There are several theories employed to interpret the absorption performance such as interference, effective medium, and equivalent circuit theory\textsuperscript{\cite{18}}. In particular, the condition of critical coupling in terms of systematic losses based on temporal coupled-mode theory (CMT)\textsuperscript{\cite{19}} enables a versatile platform to understand the perfect absorption when nonradiative loss matches radiative loss of the photonic system. Therefore, tailoring systematic loss and interpreting the mode properties play the key roles in engineering the absorption performance. 

Nonradiative loss roots in materials and fabrication defects and will be an invariable once the configuration of absorber is chosen, and carefully tuning radiative loss of the photonic cavity is the only pathway to engineer the coupling regime. A versatile control of radiative loss from zero to a finite value is thus essential in this process. Bound states in the continuum (BICs) have been found to have zero radiative loss and collapse to quasi-BICs with a controllable radiative loss by introducing structural asymmetry whose radiative loss shows a negative quadratic dependence on asymmetry\textsuperscript{\cite{20}}. BICs are intrinsic wave states that exist in the radiation continuum but perfectly confined within the resonator without coupling to free space, thus stopping the radiation. BIC has become a widely adopted mechanism to boost light-matter interactions for applications in sensing and lasing by employing its capability to avoid radiative loss\textsuperscript{\cite{21, 22}}. For applications of BIC theory in electromagnetic absorbers, narrow linewidth absorption at a single frequency point was commonly discussed which merely focused on the most prominent capability of radiative loss suppression while commonly ignoring the practical demands such as angular tolerance, engineerable operation bandwidth and film flexibility\textsuperscript{\cite{23,24,25}}. Currently, a series of significant features have been reported for BICs by tailoring their topological properties in momentum space, for example, quality factor robustness to defects\textsuperscript{\cite{26}}, directional radiation\textsuperscript{\cite{27}}, and intrinsic chirality\textsuperscript{\cite{28}}. However, most studies concentrated on geometric parameters of resonator itself whose accurate trimming adjusts the merging and separation of topological charges in momentum space but usually in an extreme condition and/or with unconventional fabrication process. Another abundant freedom to study BICs is on the lattice dimension which would enable a quite flexible channel to tune BIC and alleviate the rigid requirements of geometric parameters\textsuperscript{\cite{29}}. Therefore, a comprehensive exploration of the versatile radiative properties of BICs would enable a systematic path for electromagnetic absorbers with on-demand performance based on CMT such as large absorption coefficient, great angular stability, and controllable bandwidth.

In this work, we numerically and experimentally investigate the versatile radiative properties of BICs from the perspective of lattice symmetry, and demonstrate perfect absorption empowered by them allowing for the tailoring of absorption coefficient, bandwidth, and field of view. The absorption properties are analyzed using CMT, revealing the dominant role of BICs, whose radiation features are studied through multipolar and topological analysis in real and momentum spaces, respectively. By leveraging the unique radiation properties of lattice-symmetry-broken induced BICs, we showcase the stable perfect absorption of a thin and flexible metasurface across a wide field of view covering $\pm55^\circ$. Additionally, we establish a flexible pathway to accessing multiple BICs within an interested frequency range through the band-folding process, leading to the feasibility of achieving multi-band perfection absorption by tailoring their radiation properties. Our systematic study highlights the significant potential of BICs for optical applications and underscores their potential benefits for optoelectronic devices, particularly in the realm of terahertz detectors, sensors, and next-generation wireless communications.

\section{Results}
\subsection{Theoretical model}
According to CMT\textsuperscript{\cite{30}}, for a single-mode one-port system (Fig. \ref{Fig1}(a)), absorption spectrum is expressed as 
\begin{equation}\label{E1}
{\begin{split}
	A(\omega ) = \frac{{4{\gamma _a}{\gamma _r}}}{{{{(\omega  - {\omega _0})}^2} + {{({\gamma _a} + {\gamma _r})}^2}}}
	\end{split}}
\end{equation}
where $\omega$ and $\omega_0$ are angular frequencies of incident light and resonant frequency of the system, respectively. $\gamma_r$ and $\gamma_a$ represent loss rates originated from radiative and nonradiative contributions (e.g., material absorption), respectively. The absorption properties could be freely tailored by manipulating the two loss rates, and a perfect absorption at the resonant frequency $\omega_0$ occurs at the critical coupling condition when $\gamma_r$ is equal to $\gamma_a$\textsuperscript{\cite{19}}. In terms of absorption bandwidth, full width at the half of the maximum (FWHM) is characterized by $\Delta \omega  = {\omega _0}/{Q_t}$ where ${Q_t} = {\rm{ }}{\omega _0}/2\left( {{\gamma _r} + {\gamma _a}} \right)$ is total quality factor of the absorption spectrum\textsuperscript{\cite{31,32}}. Once the materials and configuration are settled, $\gamma_a$ could be considered as a constant, and radiative properties ($\gamma_r$) of the resonant mode are the only variable that finally determines the spectral features of absorbers.

Fortunately, BIC enables a rich physics to engineer the radiative properties that could be adopted to meet the specific requirements of absorbers. Taking the type of symmetry-protected BICs as a typical example, we explore the radiative properties caused by perturbations from the dimension of \textit{lattice} instead of \textit{resonators} where the dependence of radiation on symmetry breaking of resonators in a uniform lattice (U-BIC) has been clearly understood\textsuperscript{\cite{29,33,34}}. We choose the classical DSRR to form a unit cell in a square lattice, and the mode features are calculated as the circle points shown in Fig. \ref{Fig1}(c). A perturbation between the neighboring two DSRRs along $x$-direction will result in a modification of the square lattice to a rectangular one with doubled period (Fig. \ref{Fig1}(b)), and the corresponding first Brillouin zone (BZ) shrinks to half of the original one of square lattice along $k_x$ in momentum space. The high-symmetry point (X point) of original BZ is folded to $\Gamma$ point of BZ of rectangular lattice. In the band folding process, we focus on two guided modes at X point in the bands of the lowest frequencies (termed as GM I at lower frequency and GM II at higher frequency). They originally locate below the light line in BZ of square lattice, and are folded to $\Gamma$ point in the rectangular lattice above light line residing in the radiation continuum which thus possess the properties of BICs (G-BIC I and G-BIC II). The band structures and their radiation properties (characterized by radiative ${Q_r}$, ${Q_r} = {\rm{ }}{\omega _0}/2{\gamma _r}$) of G-BIC I, G-BIC II, and U-BIC are depicted in Fig. \ref{Fig1}(c), revealing no coupling to the continuum at $\Gamma$ point. Specifically, G-BIC I and G-BIC II reveal no leakage to the continuum at various wavevectors in contrast to U-BIC and possess a stable radiative property at different incident angles. 

The radiative stability of G-BICs will remain when the channel of coupling to the continuum is switched on through the introduction of perturbations that break the symmetry in the rectangular lattice\textsuperscript{\cite{35,36}}. Two types of perturbations are discussed which open the radiation channels of G-BIC I and G-BIC II, respectively: (i) near-field coupling strength by perturbing the distance of the neighboring unit cells (asymmetry degree is defined as ${\alpha_D}\, = \,\left( {{d_0} - d} \right)/{p_x} \times 100\%$) and (ii) resonator area by perturbing the arm length of the alternating square DSRRs (asymmetry degree is defined as ${\alpha _L}\, = \,\left( {{l_2} - l} \right)/l \times 100\%$). As illustrated in Fig. \ref{Fig1}(d), G-BIC I deteriorates to a radiative mode (G-QBIC I) with a finite $Q_r$ due to perturbation of coupling strength while G-BIC II reveals a typical wavevector-dependent radiation property (i.e., $Q_r \propto k_x^{-2}$) similar as the original U-BIC. Inherited from the intrinsic radiation stability, the radiation of G-QBIC I still exhibits independent on wavevectors $k_x$ whose strength could be manipulated by tuning $\alpha_D$. Similarly, the deterioration of G-BIC II to G-QBIC II is observed with stable radiative quality factors by perturbing the resonator area while G-BIC I reveals a wavevector-dependent radiation with properties of U-BIC unaffected as shown in Fig. \ref{Fig1}(e). In addition, the overall radiative properties of the three modes still maintain with various $k_y$ while keeping the incident polarization along $y$-axis (see Fig. S3 in supplementary information for details).

\begin{figure*}
	\centerline{\includegraphics[width=0.92\textwidth]{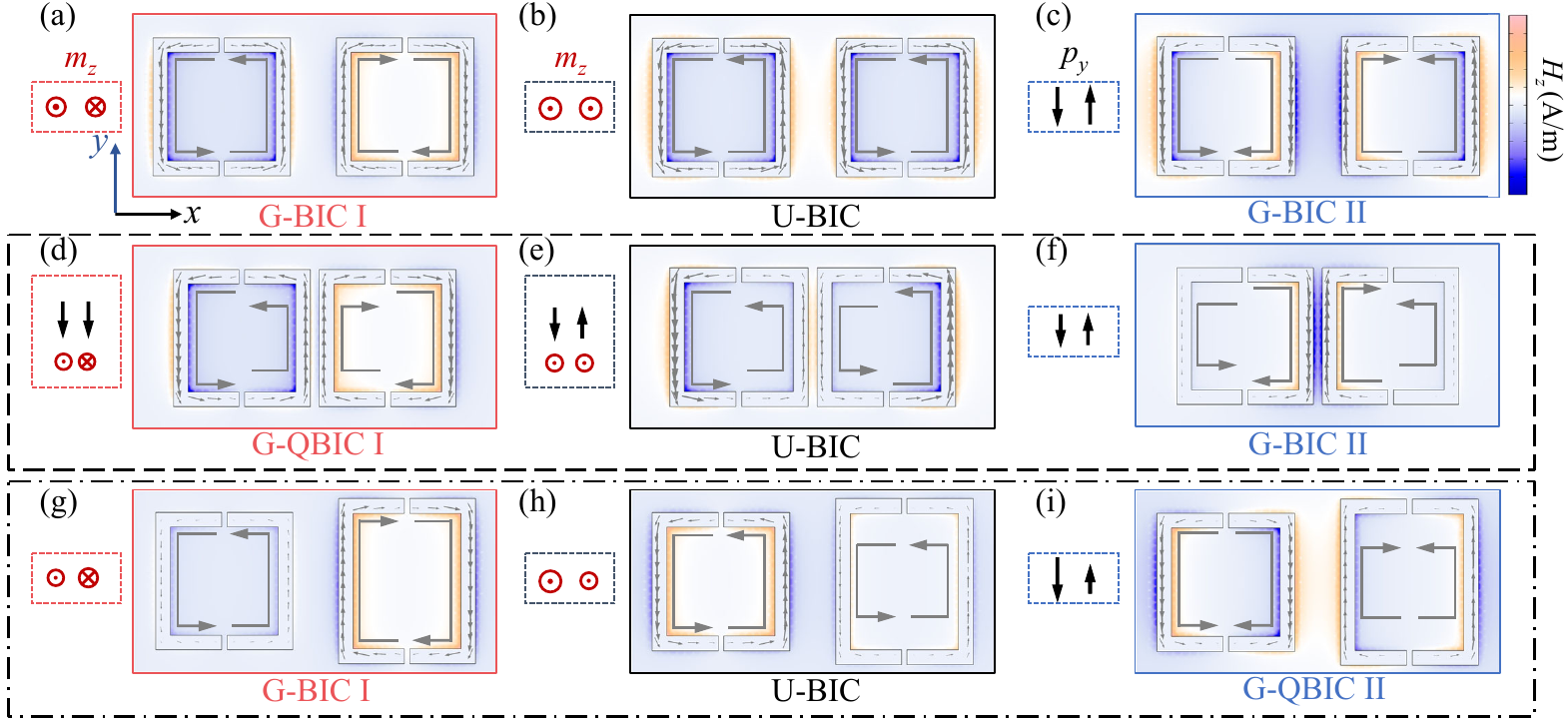}}
	\caption{{\bf Surface current distributions and magnetic fields of the three modes with and without perturbation.} (a)-(c) Surface current distributions (grey solid lines) and $z$ component of magnetic field ($H_z$) for G-BIC I, U-BIC, and G-BIC II without lattice perturbation. The insets summarize the dominant collective multipoles at each mode. Similar analysis is shown with lattice perturbations via symmetry breaking from coupling strength ((d)-(f)) and alternant resonator area ((g)-(i)).}
	\label{Fig2}
\end{figure*}

We interpret the radiation properties under the two types of perturbations via visualizing the surface current and magnetic field ($H_z$) distributions of these modes at $\Gamma$ point. The eigenmodes of G-BIC I, U-BIC and G-BIC II are calculated without perturbation in Figs. \ref{Fig2}(a)-\ref{Fig2}(c). Overall, every branch of DSRRs supports an oscillating current at the lowest order modes that forms a radiating dipole. However, the collective behavior resulting from relative phases of these dipoles determines far-field radiation. The cease of radiation of U-BIC originates from the non-radiative magnetic dipoles ($m_z$) induced by the out-of-phase currents in each DSRR. Regarding the band-folded G-BICs, far-field radiation is similarly ceased due to the out-of-phase currents with a certain symmetry, but in different fashions from U-BIC. The low-frequency G-BIC I also induces magnetic dipoles ($m_z$) but is out-of-phase between the neighboring unit cells in contrast to U-BIC, while high-frequency G-BIC II induces a collective out-of-phase (antiparallel) electric dipoles $p_y$ in each DSRR which cancels with each other in the far field. Different types of symmetry breaking will open the specific radiation channels according to their near-field current distributions which determines the multipolar components. For example, $p_y$ component dominates the radiation when the symmetry of coupling strength is perturbed for G-BIC I, which deteriorates to a radiative G-QBIC I while non-radiating features of U-BIC and G-BIC II are left intact as illustrated in Figs. \ref{Fig2}(d)-\ref{Fig2}(f). Similarly, variation of nearest-neighboring resonator area leads to a net $p_y$ component of G-BIC II while non-radiating properties of G-BIC I and U-BIC are left intact (Figs. \ref{Fig2}(g)-\ref{Fig2}(h)). According to above analysis, the strength of net radiative $p_y$ component will be determined by the asymmetry degrees ($\alpha_D$ and $\alpha_L$) under corresponding types of perturbations that would lead to a controllable resonance bandwidth for perfect absorbers.

\subsection{Radiative properties of BICs in MIM metasurfaces}
The unambiguous understanding of radiating features in these BIC modes would guide the design of on-demand absorbers. Here, we employ the typical metal-insulator-metal (MIM) configuration where the patterned metallic layer enables the access to tailor the radiative loss and the bottom metallic layer functions as a reflective mirror that prevents transmission with its thickness exceeding the skin depth of terahertz waves\textsuperscript{\cite{37}}. Since only reflection channel is allowed, absorption $A\left( \omega  \right)$ is simply related to reflection $R\left( \omega  \right)$ by $A\left( \omega  \right) = {\rm{ }}1{\rm{ }} - R\left( \omega  \right)$. The middle insulating layer is designed with a 50 $\mu$m-thick polyimide (PI, $\varepsilon  = {\rm{ }}2.96{\rm{ }} + {\rm{ }}0.10i$) functioning as a flexible spacer. 200 nm-thick aluminum (Al, $\sigma  = {\rm{ }}3.56{\rm{ }} \times {\rm{ }}{10^7}$ S/m) is adopted for the two metallic layers, and geometric dimensions of DSRR are period $p$ = 100 $\mu$m, length $l$ = 75 $\mu$m, width $w$ = 8 $\mu$m, and gap $g$ =3 $\mu$m. The samples were fabricated by using conventional photolithography and thermal evaporation, and reflection spectra were measured with terahertz time domain spectroscopy system (THz-TDS, see Materials and methods for fabrications and measurements). The samples are free from a rigid substrate and is thus flexible with an ultrathin film (~50 $\mu$m in total). The optical image of the flexible MIM absorber is shown in Fig. \ref{Fig4}(a).

\begin{figure*}
	\centerline{\includegraphics[width=0.82\textwidth]{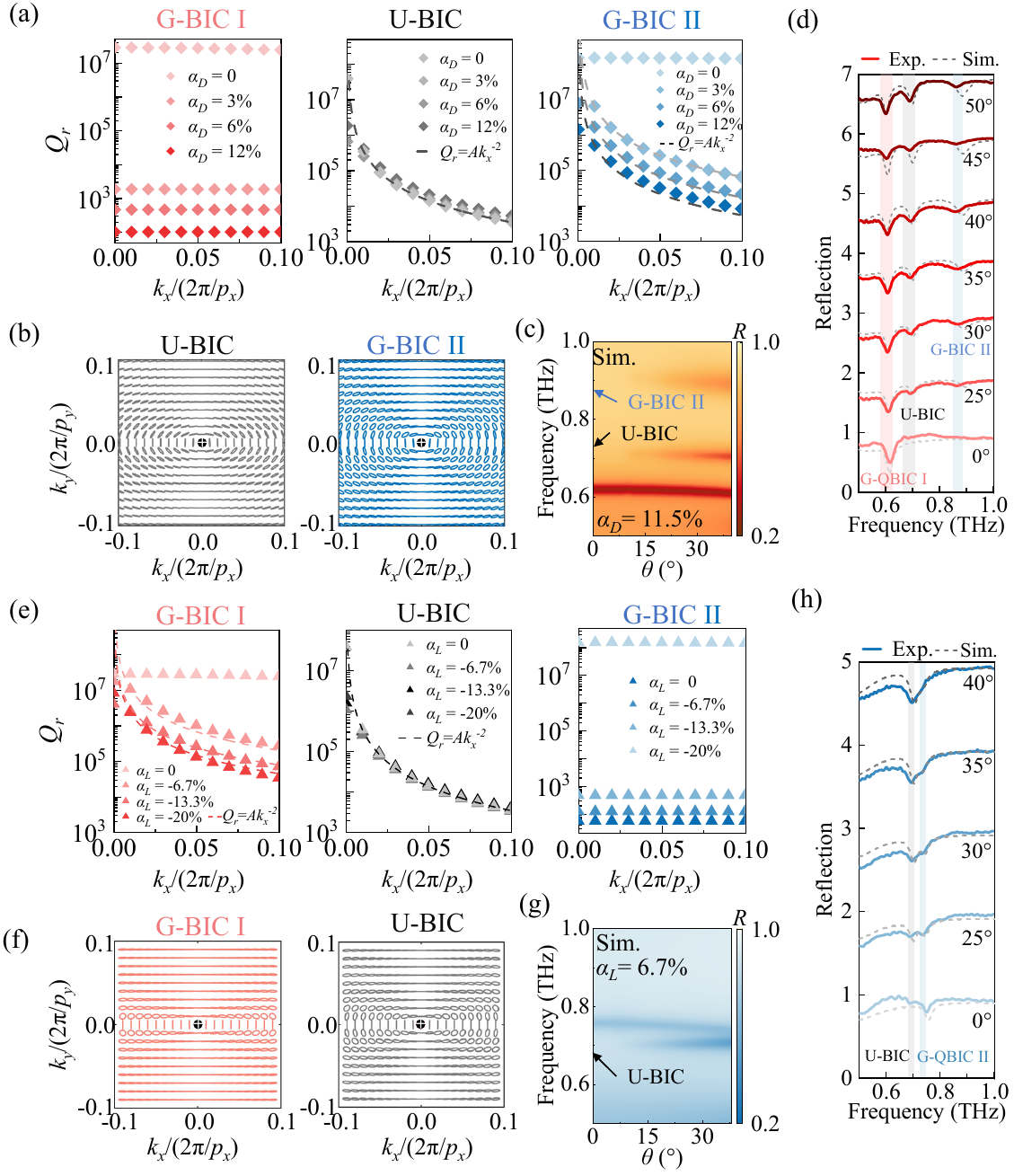}}
	\caption{{\bf Demonstration of mode properties in simulations and experiments in momentum space.} Radiative quality factors of G-BIC I (red), U-BIC (grey) and G-BIC II (blue) versus wavevector at various asymmetry degrees by breaking (a) coupling strength symmetry $\alpha_D$ and (e) resonator area symmetry $\alpha_L$. Dashed lines are fitting curves with $Q_r \propto k_x^{-2}$. (b) The far-field polarization maps of U-BIC (black) and G-BIC II (blue) in the momentum space with $\alpha_D$ = 11.5\% ($d$ = 2 $\mu$m). (c) Simulated angular-resolved reflection spectra, and (d) experimental (red solid lines) and simulated (grey dashed lines) reflection spectra at different TE-polarized incident angles ranging from $0^\circ$ to $50^\circ$. (f) The far-field polarization maps of G-BIC I (red) and U-BIC (black) in momentum space with $\alpha_L$= 6.7\% ($l_2$ = 80 $\mu$m). (g) Simulated reflection spectra at various incidence angles, and (h) experimental (blue solid lines) and simulated (grey dashed lines) reflection spectra with incident angles ranging from $0^\circ$ to $40^\circ$.}
	\label{Fig3}
\end{figure*}
\begin{figure*}
	\centerline{\includegraphics[width=0.96\textwidth]{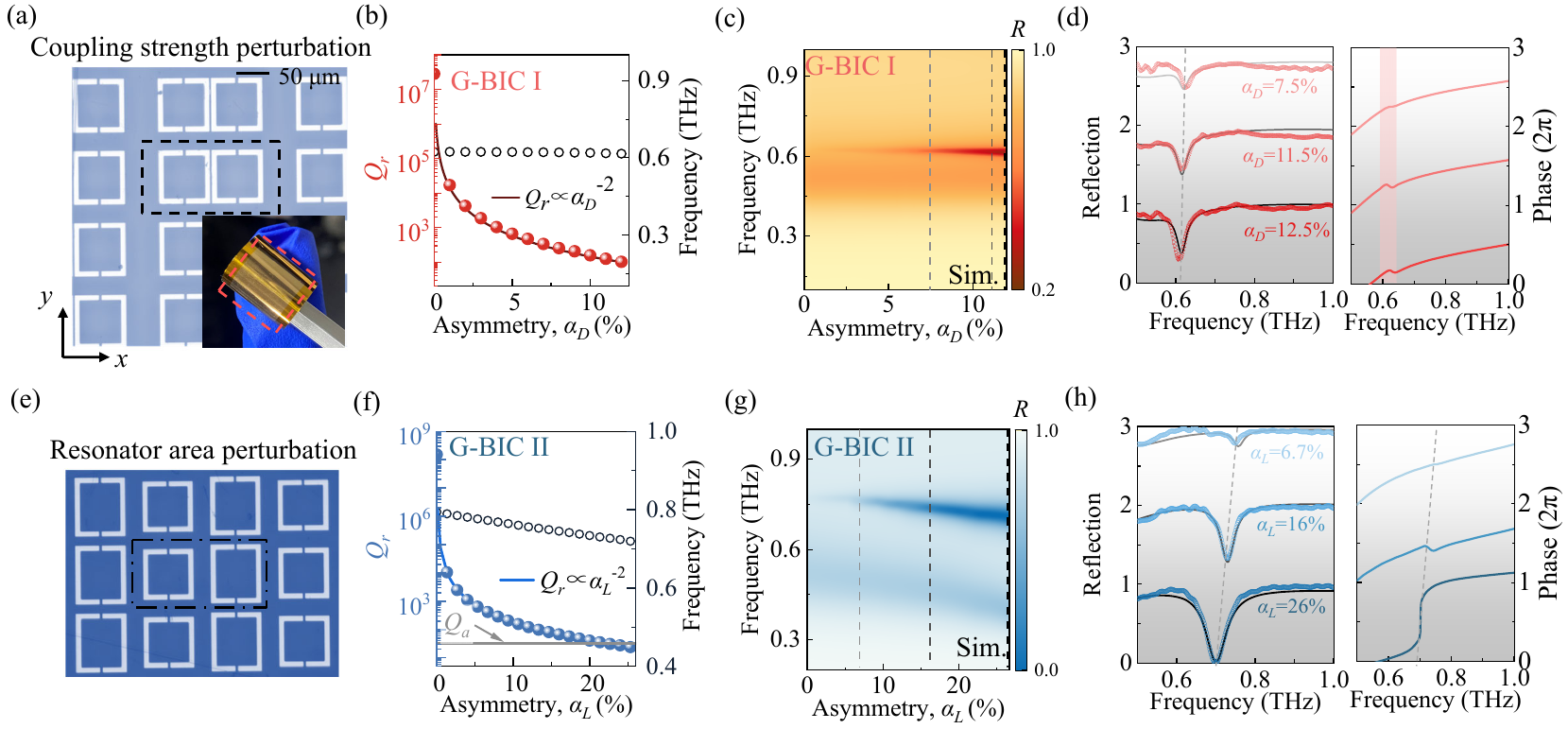}}
	\caption{{\bf Demonstration of mode properties in simulations and experiments in geometric parameter space.} (a) and (e) Microscopic images of samples by breaking lattice symmetry via disturbing the coupling distance ($d$ = 2 $\mu$m, $\alpha_D$ = 11.5\%) and resonator area ($l_2$ = 80 $\mu$m, $\alpha_L$ = 6.7\%). The inset is the optical image of a fabricated flexible sample. (b) and (f) Radiative $Q_r$ (solid dots) and central frequencies (black circles) of G-BIC I and G-BIC II by tuning $\alpha_D$ and $\alpha_L$, respectively. The radiative quality factors are fitted by $Q_r \propto \alpha_D^{-2}$ and $Q_r \propto \alpha_L^{-2}$, and the grey solid line indicates the estimated $Q_a$. (c) and (g) Evolutions of reflection intensity spectra for G-BIC I and G-BIC II versus $\alpha_D$ and $\alpha_L$ with TE-polarized incidence, respectively. (d) and (h) Experimental (red/blue dashed lines) and simulated (gray solid lines) reflection intensity and phase spectra at different asymmetry degrees.}
	\label{Fig4}
\end{figure*}
The radiation properties of BICs discussed above will remain by the patterned metasurface layer in the MIM cavity. Under the two types of perturbations on the patterned metasurface layer, we probe the radiation features from the reflection spectra as shown in Fig. \ref{Fig3}. In the scenario of coupling strength perturbation of the lattice ($d$ = 2 $\mu$m, $\alpha_D$ = 11.5\%), we envision the deterioration of G-BIC I to G-QBIC I while G-BIC II and U-BIC are left intact according to the mode analysis. It is evidenced by the finite value of $Q_r$ for G-QBIC I, and the inverse quadratic dependence of $Q_r$ of U-BIC and G-BIC II on $k_x$ ($Q_r \propto k_x^{-2}$, Fig. \ref{Fig3}(a)). Regarding U-BIC and G-BIC II, the vortex centers of polarization vectors in momentum space reveal undefined polarization state at $\Gamma$ point (Fig. \ref{Fig3}(b)) showing a topological charge of +1, which further indicates non-radiative nature of the two BIC modes\textsuperscript{\cite{38}}. Despite the similarity of non-radiative nature, U-BIC and G-BIC II intrinsically possess different dependence on asymmetry degree where G-BIC II reveals a clear modulation of radiation at various values of $\alpha_D$ similar to that of G-QBIC I ($Q_r \propto \alpha_D^{-2}$, will discuss in Fig. \ref{Fig4}). With the increase of degree of coupling strength perturbation, the radiative $Q_r$ of band-folded G-BIC II will decrease, while the decay rate is insensitive to the specific type of perturbation in the original U-BIC. Such an interesting feature of G-BIC II could enable a path toward engineering the quality factor dispersion benefiting angle-multiplexed multispectral applications\textsuperscript{\cite{39}}. 

We observe the far-field radiation properties through angle-resolved simulations and experiments. The resonance linewidths of G-BIC II and U-BIC gradually diminish and finally vanish as the incidence angle shifts back to normal (Fig. \ref{Fig3}(c)) while G-QBIC I exhibits almost constant resonant properties as the incident angle varies. In experiments, terahertz reflection spectra are obtained with TE polarization where terahertz time-domain spectroscopy system is reconfigured for angular-resolved measurements. The normalized reflection intensity spectra in experiments ranging from $0^\circ$ to $50^\circ$ reproduce the theoretical and simulated results (grey dashed lines for comparison). Three resonances dips (G-QBIC I at 0.61 THz, U-BIC at 0.69 THz and G-BIC II at 0.87 THz) are visualized in the spectra under tilted incidence while only G-QBIC I remains at normal incidence with stable central frequency and resonant linewidth consistent with our analysis. 

Similar analysis, simulations and measurements are carried out for the scenario ii where resonator area is perturbed with length of alternant DSRR adjusted to $l_2$ = 80 $\mu$m ($\alpha_L$= 6.7\%) along $x$ axis. Here, the low-frequency G-BIC I reveals a +1 topological charge in momentum space as well as an inverse quadratic relationship between $Q_r$ and $k_x$ ($Q_r \propto k_x^{-2}$) together with U-BIC (Figs. \ref{Fig3}(e) and \ref{Fig3}(f)). It is noted that $Q_r$ of G-BIC I keeps holding a large value at large wavevectors whose linewidth exceeds the resolution of apparatus, and it is thus hardly captured in far-field reflection spectra (Figs. \ref{Fig3}(g) and \ref{Fig3}(h)). Leakage of U-BIC at 0.69 THz and G-QBIC II at 0.75 THz are observed in the spectra at tilted angles of incidence, and the resonance of G-QBIC II persists at normal incidence in the angular-resolved reflection spectra (Fig. \ref{Fig3}(h)).

\begin{figure*}
	\centerline{\includegraphics[width=0.78\textwidth]{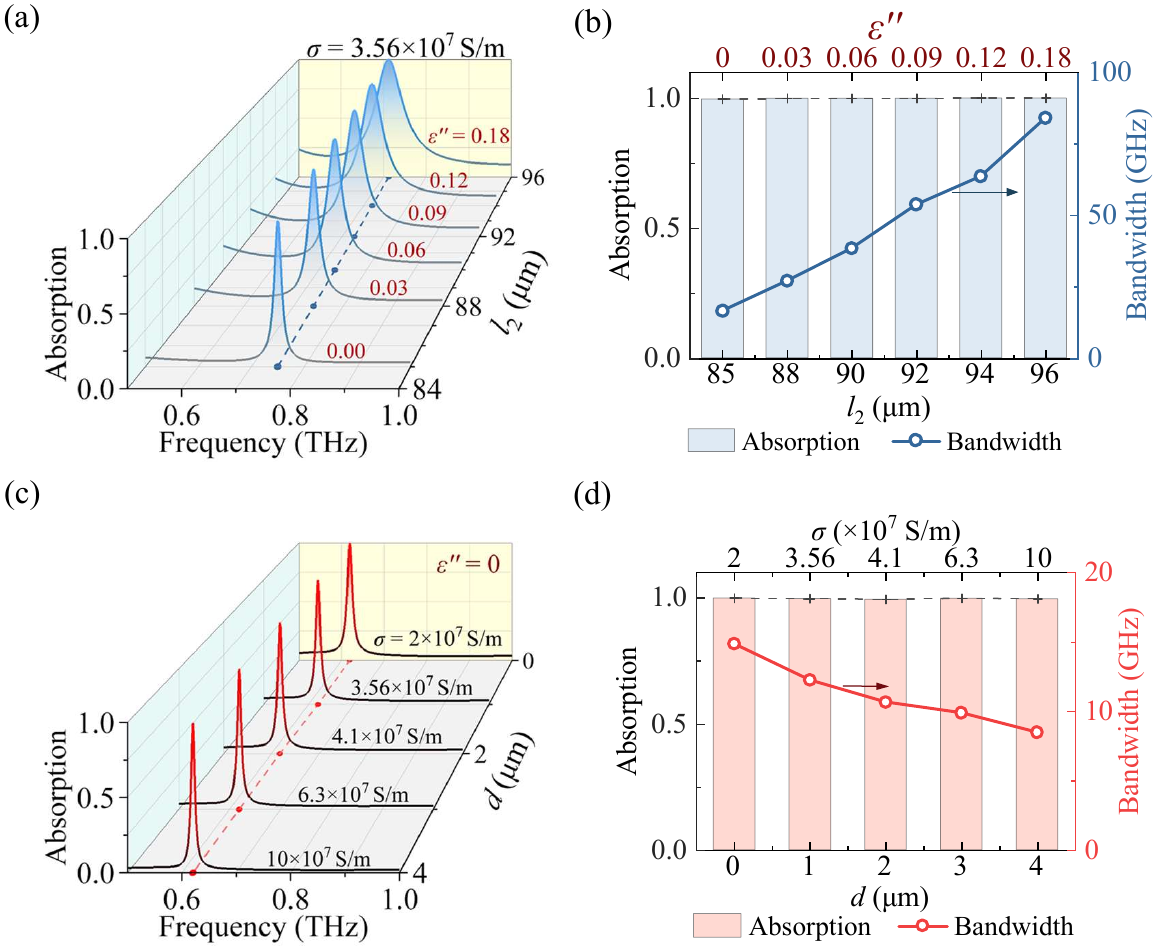}}
	\caption{{\bf Bandwidth manipulation of perfect absorption.} (a) Evolution of absorption bandwidth by adjusting $\alpha_L$ with G-BIC II while keeping perfect absorption for insulating layers with various imaginary parts of permittivity, and (b) the extracted bandwidth. (c) Evolution of absorption bandwidth by adjusting $\alpha_D$ with G-BIC I while keeping perfect absorption for metallic parts with various conductivities, and (d) the extracted bandwidth.}
	\label{Fig5}
\end{figure*}

An unambiguous interpretation of radiation properties in these BIC-empowered modes provides a versatile access to engineer the absorption spectrum in metasurface absorbers. We then move to probe the conditions of critical coupling by adjusting the asymmetry degrees at normal incidence ($\Gamma$ point in the momentum space) for G-BIC I and G-BIC II, respectively. In the scenario of coupling symmetry perturbation, we concentrate on G-BIC I whose resonance frequency reveals negligible dependence on asymmetry degree while $Q_r$ possesses inverse quadratic dependence on asymmetry ($Q_r \propto \alpha_D^{-2}$) as shown in Figs. \ref{Fig4}(b) and \ref{Fig4}(c). Samples with three typical values of coupling asymmetry are fabricated to perform measurements, and the corresponding intensity and phase of reflection spectra are shown in Fig. \ref{Fig4}(d) with good agreement between experiments and simulations. As revealed by the resonance depth and corresponding phase spectra, all the three scenarios operate in overdamped regime\textsuperscript{\cite{19}}, and gradually approach the critical coupling condition by increasing the asymmetry degree accompanying with a decrease of their total quality factors $Q_t$ as predicted. Limited by the geometric distance between the neighboring resonators, it is infeasible to further increase the asymmetry degree in practice.

\begin{figure*}
	\centerline{\includegraphics[width=0.86\textwidth]{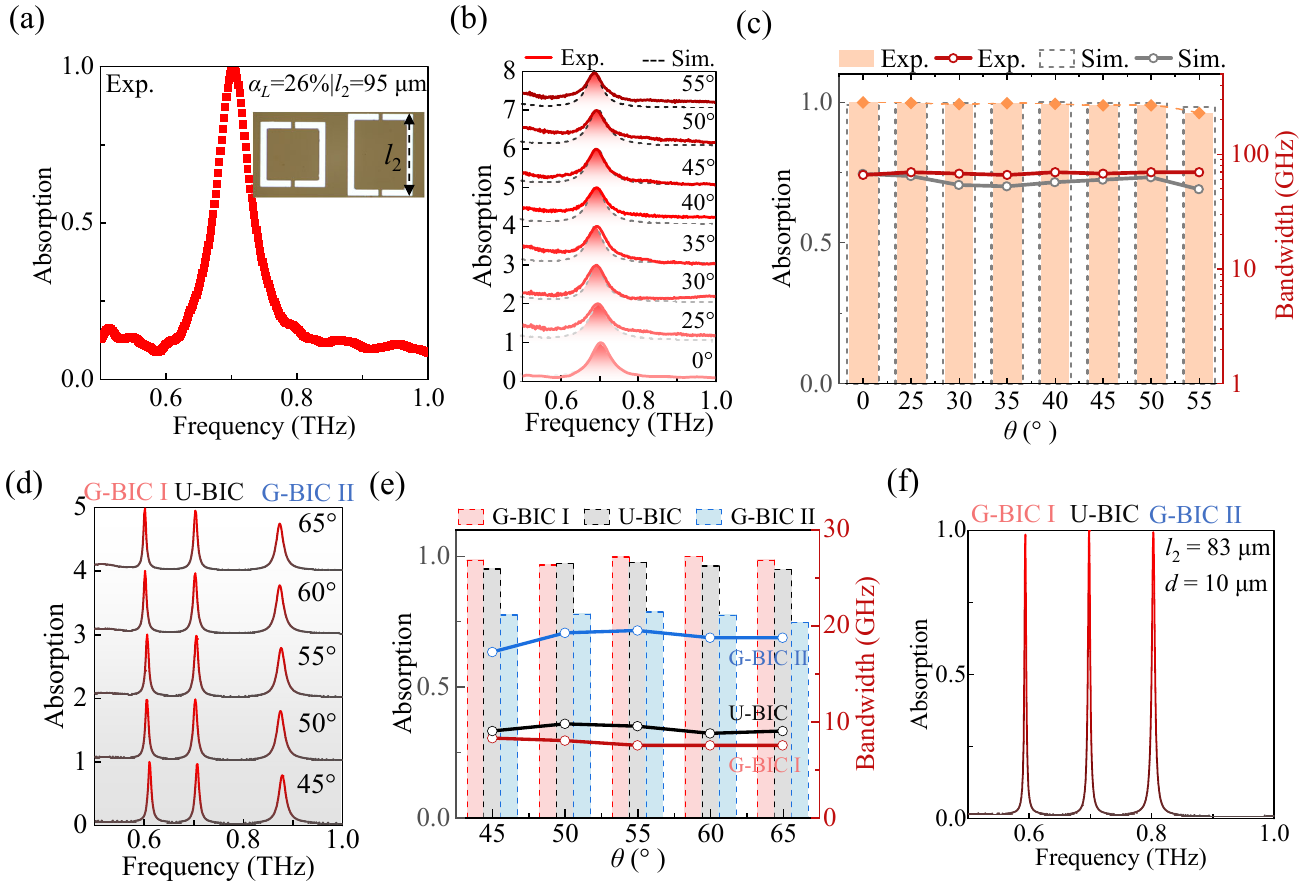}}
	\caption{{\bf Incident-angle stability of perfect absorption covering a large field of view and multi-band perfect absorption.} (a) Experimental perfect absorption spectrum with G-BIC II and $\alpha_L$ = 26\% at normal incidence. The inset is microscopic image of the fabricated sample. (b) Experimental (red solid lines) and simulated (grey dashed lines) absorption spectra at different incident angles. (c) The extracted absorption coefficient (bar chart) and bandwidth (solid dotted line) in experiments and simulations. (d) Dual-band perfect absorption with the freedoms of incident angle and asymmetry degree $\alpha_D$ ($\alpha_D$ = 11\%).  (e) The extracted absorption coefficient (bar chart) and bandwidth (solid dotted line) empowered by G-BIC I (red), U-BIC (grey) and G-BIC II (blue). (f) Triple-band perfect absorption with geometric freedoms of both asymmetry degrees at normal incidence.}
	\label{Fig6}
\end{figure*}

A clear access to the critical coupling condition is realized in the scenario of resonator area perturbation which enables a larger value of asymmetry degree so that radiative loss could approach the value of intrinsic non-radiative loss. Here, G-BIC II is excited at higher frequency whose resonance frequency shows a slight shift with $\alpha_L$ due to the change in geometry size of resonators when adjusting $\alpha_L$, and $Q_r$ still follows the rule of $Q_r \propto \alpha_L^{-2}$ (Figs. \ref{Fig4}(f) and \ref{Fig4}(g)). According to experimental spectra, the critical condition is expected to occur around $\alpha_L$ = 26\% where reflection intensity equal to zero and a $2\pi$ phase coverage are presented in Fig. \ref{Fig4}(h) as a typical signature of underdamped regime around critical coupling condition\textsuperscript{\cite{40,41}}. We further optimize the asymmetry degree and estimate the non-radiative quality factor around 26.3 at 0.7 THz based on CMT as the grey solid line shown in Fig. \ref{Fig4}(f) which determines the final bandwidth of perfect absorption ($Q_t$ = $Q_a/2$ = 13.15) in the critical coupling condition (see supplementary information).

\subsection{Performance of terahertz perfect absorbers}
The above discussion indicates an access to manipulate the bandwidth of a perfect absorber which is crucial for applications in different scenarios such as sensing and detection. As we conclude from CMT, the total quality factor and bandwidth of perfect absorption will be determined by the inevitable non-radiative loss (major contribution from material loss) and the subsequently matched radiative loss which is a customizable parameter. As a proof of demonstration, we adjust the imaginary part of refractive index ($\varepsilon''$) of spacer material and conductivity ($\sigma$) of metallic layer, respectively, to tune the non-radiative loss; and then balance the system to the critical coupling condition by introducing different asymmetry degrees employing the G-BICs. We are able to obtain various bandwidths ranging from 16 GHz to 84 GHz by adjusting $\varepsilon''$ and asymmetry degree while conserving perfect absorption at near 0.70 THz with G-BIC II (Figs. \ref{Fig5}(a) and \ref{Fig5}(b)). Although the tuning range of radiative loss of G-BIC I is limited due to the restricted coverage of asymmetry degree, it could be adopted for perfect absorption with sharp and frequency-stable spectra in a low-loss system. By simply adjusting the conductivity of metallic layer without considering loss from spacer (approximated by high-resistivity silicon or polymethylpentene, TPX), sharp perfect absorption spectra are obtained with a stable resonance frequency at 0.625 THz and much narrower linewidth as shown in Figs. \ref{Fig5}(c) and \ref{Fig5}(d). 

When the critical coupling condition is reached, the stable radiation feature at various wavevectors of G-BICs guarantees the perfect absorption with unaltered bandwidth covering a large field of view. Taking G-BIC II as a typical example, we characterized the angular-resolved spectral responses of the sample with asymmetry degree $\alpha_L$ = 26\%. Nearly-perfect absorption with coefficient $A\left( {{\omega _0}} \right){\rm{ }} > {\rm{ }}99\%$ is achieved at $\omega_0$ = 0.70 THz with a quality factor $Q_t$ = 11.2 and bandwidth of 62.5 GHz at normal incidence (Fig. \ref{Fig6}(a)). Absorption spectra at a series of incident angles from $0^\circ$ to $55^\circ$ are captured and shown in Fig. \ref{Fig6}(b) which match well with simulations. The spectral performance is characterized by absorption coefficient and bandwidth as summarized in Fig. \ref{Fig6}(c) exhibiting an excellent stability in the field of view covering the entire range of $55^\circ$. The angular-insensitive feature would especially benefit applications in energy harvesting and aircraft stealth together with its low-cost fabrication and flexible capability of close adherence to uneven surfaces.

We have demonstrated the capability of BIC-empowered modes for customizing the metasurface absorber with desired bandwidth, large angle tolerance and perfect absorption coefficient in an ultrathin and flexible film with a single mode. In some circumstances, multiband absorption would be demanded for ultrabroad band or multi-spectral applications\textsuperscript{\cite{42}}. As we have demonstrated in Fig. \ref{Fig3}, all the three modes will be excited at nonzero wavevectors $k_x$ by breaking lattice symmetry via disturbing either coupling distance or alternant resonator area. It would be feasible to simultaneously meet the critical coupling condition for two modes with the two parameter freedoms -- lattice symmetry and incident angles. In the scenario of coupling strength symmetry breaking at $\alpha_D$ = 11\% ($d$ = 3 $\mu$m), dual-band near-perfect absorption with stable bandwidth originated from G-BIC I and U-BIC is observed at incidence angles ranging from $45^\circ$ to $60^\circ$ while the one originated from G-BIC II reveals a relatively faint absorption due to mismatch of losses as shown in Figs. \ref{Fig6}(d) and \ref{Fig6}(e). Here, the conductivity of metal is set as $\sigma  = {\rm{ }}6.3 \times {10^7}$ S/m (e.g., silver or gold) with a lossfree spacer layer. In terms of triple band perfect absorption, the intrinsically larger quality factor of G-BIC II would require a lower non-radiative loss to reach the critical coupling condition that could be achieved with a larger conductivity of metals\textsuperscript{\cite{43,44}} (see Fig. S5 in supplementary information for details). 

Moreover, simultaneously breaking both lattice symmetries via disturbing coupling distance as well as resonator area will make all the three modes leaky at normal incidence (see Fig. S6 in supplementary information for details)\textsuperscript{\cite{36}}. It would thus be feasible to access triple-band perfect absorption at normal incidence via tailoring non-radiative loss and parameters of symmetry breaking. As illustrated in Fig. \ref{Fig6}(f), triple-band perfect absorption originated from G-BIC I, U-BIC, and G-BIC II at 0.59 THz, 0.70 THz and 0.80 THz are observed at normal incidence with $\sigma  = {\rm{ }}6 \times {10^8}$ S/m, $d$ = 10 $\mu$m ($\alpha_D$ = 7.5\%) and $l_2$ = 83 $\mu$m ($\alpha_L$ = 10.7\%). Compared to traditional strategies for multi-band absorption, such as employing multiple area-gradient resonators in a supercell\textsuperscript{\cite{15,16}} and stacking multi-layer metasurface structures\textsuperscript{\cite{45}}, the strategy based on CMT and BICs not only provides a comprehensive interpretation of mode properties but also alleviates the fabrication complexity with a single layer metasurface without extreme requirement on geometric parameters.

\section{Discussion}
In summary, our systematic study of the radiation properties of BICs in the context of lattice symmetry has enabled the development of terahertz absorbers with several key features, including tunable bandwidth, stable performance in a wide field of view covering $\pm55^\circ$, and multi-band perfect absorption. The capability of tunable bandwidth would enhance the performance of detectors by improving spectral resolution, signal to noise ratio, and sensitivity. Furthermore, the stable performance in a wide field of view guarantees the compatibility with curved surfaces together with their thin and flexible configuration, facilitating their applications in wearable devices and as invisibility layers of aircraft. By simultaneously matching three or more BIC modes at critical coupling conditions, multispectral or hyperspectral applications are envisioned for sensing and imaging purposes. Overall, our findings provide valuable insights for the development of optoelectronic devices based on metamaterials, particularly in terahertz regime, for efficient detectors and sensors, as well as their applications in communications and imaging.

\section{Materials and methods}
\noindent
{\bf Sample Fabrication.} The samples were fabricated by using traditional photolithography process on a 50 $\mu$m-thick polyimide film with an area of 15 mm $\times$ 15 mm. Initially, a 2 $\mu$m-thick photoresist (S1818) was spin-coated on the cleaned PI substrate at a speed of 4500 r/s for 60 s, followed by a pre-bake process of the film at 110 \textdegree{}C for 60 s. Subsequently, the pattern was transferred from a mask to the photoresist using ultraviolet photolithography (SUSS-MA6). The sample was then developed for 60 s to remove the unexposed photoresist, and then the patterned film was baked at 120 \textdegree{}C for 90 s. A 200 nm-thick aluminum was deposited on the patterned film by using thermal evaporation. Lift-off process of the remaining photoresist was carried out by sonicating in acetone for 5 minutes. Finally, another round of thermal deposition was carried out to deposit a 200-nm thick aluminum on the cleaned backside of the fabricated film. Optical microscopic images of fabricated samples are shown in Figs. \ref{Fig4} and \ref{Fig6}.

\vspace{\baselineskip}

\noindent
{\bf Optical measurements.} The measurements were performed using a commercially available THz-TDS and the reflection intensity spectra were calculated by $R\left( \omega  \right){\rm{ }} = {\rm{ }}{\left| {{E_S}\left( \omega  \right)/{E_R}\left( \omega  \right)} \right|^2}$, where ${E_S}\left( \omega  \right)$ and ${E_R}\left( \omega  \right)$ are the Fourier transforms of the time-domain electric fields of the samples and reference (Al reflective mirror), respectively. The setup was reconfigured for reflection measurements at normal incident mode and angle-resolved mode.

\section*{Data Availability Statement}
\noindent
Data presented in this work may be obtained from the authors upon reasonable request.

\begin{acknowledgments}
 \noindent
This work was supported by the National Natural Science Foundation of China (Award No.: 1), Guangdong Basic and Applied Basic Research Foundation (Award No.: 2023A1515011085), and Shenzhen Science and Technology Program (Award No.: JCYJ20230807093617036, 20220815151149004, JSGGKQTD20221101115656030). The authors acknowledge the assistance of SUSTech Core Research Facilities.
\end{acknowledgments}

\section*{Author Contributions}
\noindent
L. Cong supervised the project; G. Xu designed the whole samples; G. Xu and Z. Xue built the optical setup and performed the optical experiments; G. Xu fabricated the samples guided by J. Fan; G. Xu analyzed the data and prepared the main manuscript. All the authors read and commented on the manuscript.

\section*{Conflict of Interest}
\noindent
The authors declare no conflicts of interest.

\end{document}